\documentclass[aps,prapplied,showpacs,twocolumn,preprintnumbers,amsmath,amssymb,longbibliography]{revtex4-2}

\usepackage{graphicx}
\usepackage{dcolumn}
\usepackage{bm}
\usepackage{color}
\usepackage{amsmath}
\usepackage{upgreek}
\usepackage{float}
\usepackage[colorlinks,linkcolor=blue, urlcolor=blue, anchorcolor=blue, citecolor=blue]{hyperref}

\begin{document}

\title{Topologically protected generation of spatiotemporal optical vortices with nonlocal spatial-mirror-symmetry-breaking metasurface}

\author{Junyi Huang}
\affiliation{Interdisciplinary Center of Quantum Information, State Key Laboratory of Modern Optical Instrumentation, and Zhejiang Province Key Laboratory of Quantum Technology and Device, Department of Physics, Zhejiang University, Hangzhou 310027, China}

\author{Hongliang Zhang}
\affiliation{Interdisciplinary Center of Quantum Information, State Key Laboratory of Modern Optical Instrumentation, and Zhejiang Province Key Laboratory of Quantum Technology and Device, Department of Physics, Zhejiang University, Hangzhou 310027, China}

\author{Tengfeng Zhu}
\affiliation{Interdisciplinary Center of Quantum Information, State Key Laboratory of Modern Optical Instrumentation, and Zhejiang Province Key Laboratory of Quantum Technology and Device, Department of Physics, Zhejiang University, Hangzhou 310027, China}

\author{Zhichao Ruan}
\email{zhichao@zju.edu.cn}
\affiliation{Interdisciplinary Center of Quantum Information, State Key Laboratory of Modern Optical Instrumentation, and Zhejiang Province Key Laboratory of Quantum Technology and Device, Department of Physics, Zhejiang University, Hangzhou 310027, China}

\begin{abstract}
Recently nonlocal spatial-mirror-symmetry-breaking metasurfaces have been proposed to generate spatiotemporal optical vortices (STOVs), which carry transverse orbital angular momenta. Here we investigate the topological property of the STOV generator and show that spatial mirror symmetry breaking introduces a synthetic parameter dimension associated with the metasurface geometry. Furthermore, we demonstrate that there are well-defined vortices emerging with the synthetic parameter dimension, which can topologically protect the STOV generation robustly against structural perturbations and disorders. Since the vortices are always `created' or `annihilated' together in pairs of opposite charges in the ${{k}_{x}}\!-\!\omega $ domain, the total topological charge of these vortices is a conserved quantity. Our studies not only provide a new topological perspective for STOV generation but also lay a solid foundation for potential applications of STOV metasurfaces integrated with other optoelectronic devices, because of their robust immunity to fabrication defects.
\end{abstract}
\maketitle

Since the intriguing recognition that optical phase singularity \cite{nye1974dislocations} can lead to longitudinal orbital angular momentum (OAM)  \cite{yao2011orbital,franke2008advances,shen2019optical}, the studies of optical field singularities have been broadly extended over spatial dimensions. Recently, intense interest has been attracted to explore the optical vortices in the spatiotemporal domain \cite{sukhorukov2005spatio,dror2011symmetric,bliokh2012spatiotemporal, bliokh2015transverse,bliokh2021spatiotemporal, chong2020generation,hancock2019free,jhajj2017hydrodynamic, jhajj2016spatiotemporal, hancock2021second,wan2020generation, wan2021photonic, gui2021second,wang2021engineering,huang2022spatiotemporal, hancock2021mode, mazanov2021transverse, cao2022non, fang2021controlling, cao2021sculpturing}. In particular, spatiotemporal optical vortices (STOVs), as an optical pulse, can be tilted with respect to the propagation direction and exhibit transverse components of OAM unique in the spatiotemporal domain \cite{bliokh2012spatiotemporal,bliokh2015transverse,bliokh2021spatiotemporal}. With such transverse OAM, all the family members of optical angular momentum are united from both spin and orbital perspectives and with both transverse and longitudinal components.

Although transverse OAM offers a new degree of freedom and potential advances, it is a challenge to generate STOVs due to the requirement of neat optical manipulations associated with both spatial and temporal domains, simultaneously. By using pulse shapers, the spatiotemporal control method was firstly experimentally demonstrated to generate STOVs in free space with bulk optical systems \cite{gui2021second,chong2020generation,hancock2019free,jhajj2017hydrodynamic, jhajj2016spatiotemporal, hancock2021second,wan2020generation, wan2021photonic}. Alternatively, much more compact than pulse shapers, nonlocal metasurfaces were proposed to create phase singularities in the spatiotemporal domain by breaking spatial mirror symmetry \cite{wang2021engineering,huang2022spatiotemporal}. Moreover, we showed that the phase singularity can be used to realize differentiation computation in the spatiotemporal domain \cite{huang2022spatiotemporal}. Analog to the image processing of edge detection in the spatial domain \cite{zhu2017plasmonic, guo2018photonic, kwon2018nonlocal, saba2018two,cordaro2019high,zhu2019generalized, zhou2019optical, zhu2020optical, zhou2020flat,huo2020photonic, zhu2021topological, zangeneh2021analogue, vafa2021analog}, we proposed the spatiotemporal differentiation with the generated STOVs to detect the sharp changes of pulse envelopes, in both spatial and temporal dimensions \cite{huang2022spatiotemporal}.

\begin{figure}[htbp]
\centerline{\includegraphics[width=3.2in] {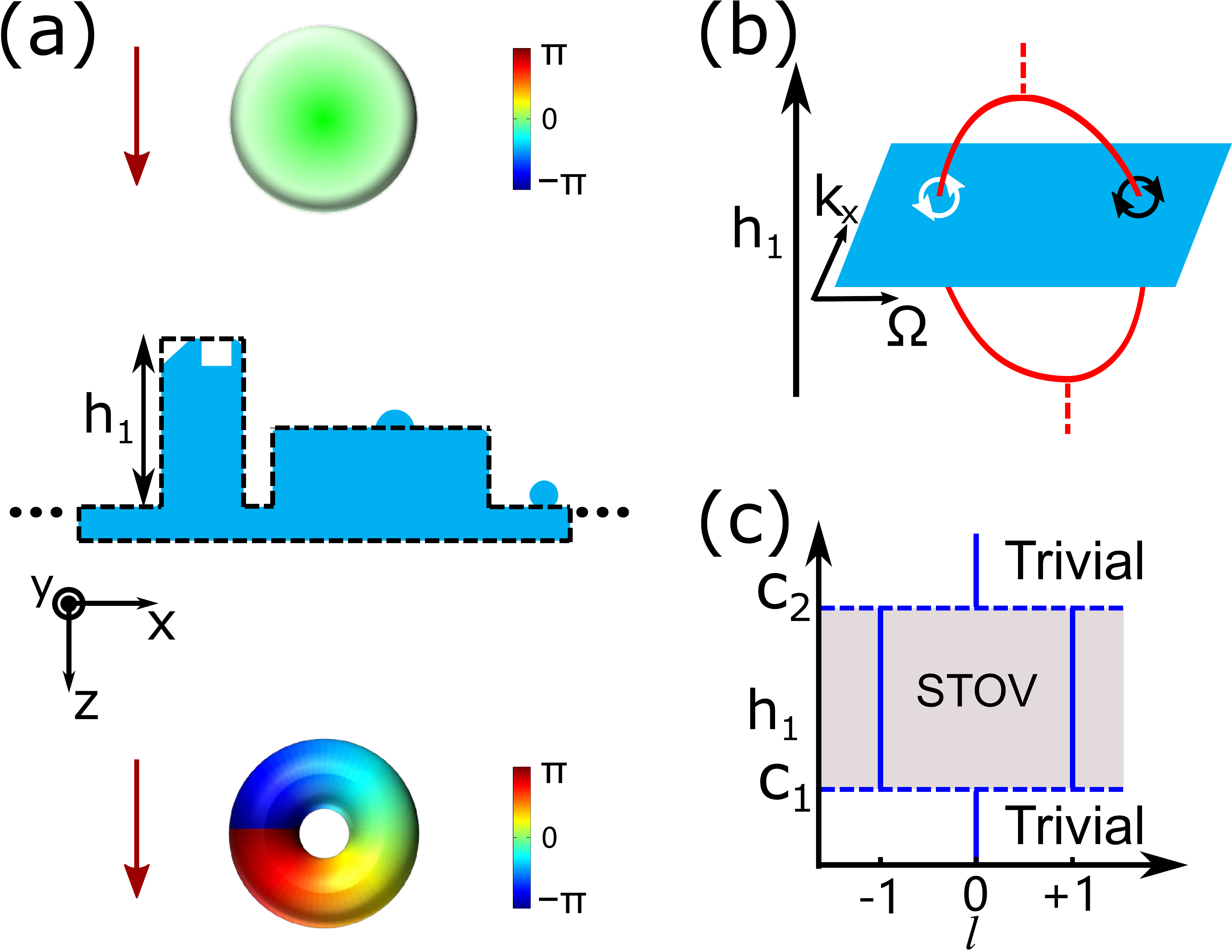}}
\caption{ (a) Schematic of topological protection for STOV generations with the metasurface, a one-dimensional periodic grating (the dashed line) perturbed with random defects along $x$ with spatial mirror symmetry breaking. The detailed geometry of the metasurface is described in Sec. I of the Supplementary Material. The transmitted pulse corresponds to a STOV carrying transverse orbital angular momentum for an incident pulse with both Gaussian envelopes in the spatial and temporal domains. (b) A closed path nodal line in the three-dimensional synthetic parameter space of ${{k}_{x}}$, $\omega $ and ${{h}_{1}}$. The white and black circles in the cross-section ${{k}_{x}}\!-\!\omega $ plane indicate the opposite topological charges, and the total topological charge of these vortices is a conserved quantity. (c) Nature of the topological protection for STOV generations: The topological phase transition emerge from the trivial cases with the topological charge $l=0$ to the nontrivial STOV one, where ${{c}_{1}}$ and ${{c}_{2}}$ correspond to the critical values of ${{h}_{1}}$.
\label{fig:1}}
\end{figure}

In this Letter, we move a step further to investigate the topological property of STOV metasurface. We show that spatial mirror symmetry breaking introduces a synthetic parameter dimension \cite{yuan2018synthetic} associated with the metasurface geometry. More importantly, the STOV generation can be topologically protected by a vortex configuration emerging with the synthetic parameter dimension, which provides robustness against structural perturbations and disorders. In fact, fabrication tolerance is of great importance for potential applications when the STOV devices are integrated with other optoelectronic devices. Originally, such robust topological protection was discovered in solid-state electronic systems \cite{bernevig2006quantum, fu2007topological, konig2007quantum, hsieh2008topological, xia2009observation, zhang2009topological} and has been extended to photonics
\cite{wang2009observation,RevModPhys.91.015006,lu2014topological,khanikaev2017two}. In particular, nontrivial photonics edge states for photonic crystals were greatly developed for robust light transmission \cite{shalaev2019robust,he2019silicon,arora2021direct,wang2020coherent,hafezi2011robust,zhao2018topological,gao2020dirac,tambasco2018quantum,barik2018topological,chen2021topologically,dutt2020single, mancini2015observation, gao2015topological, yang2018ideal, yang2019realization, wang2017optical, xie2019visualization, xia2021nonlinear}. As a key area of modern optical research, topological photonics fundamentally deepens the understanding of light interference, diffraction, and scattering. Moreover, topological protections have been explored for high performance devices against fabrication defects, such as topological delay lines \cite{hafezi2011robust}, robust lasing \cite{st2017lasing,bahari2017nonreciprocal}, infinite Q-factors resonator with bound states in continuum \cite{hsu2013observation,hsu2016bound}, polarization conversion \cite{guo2017topologically,guo2019arbitrary}, and spatial field transformation for analog computing \cite{ zhu2021topological, zangeneh2019topological}. In this context, our studies not only provide a new topological perspective for STOV generation, but also lay a solid foundation for STOV metasurfaces, because of their robust immunity to perturbations and disorders. Since the synthetic parameter dimensions can be easily extended to high dimensions by including more geometry parameters, our studies also indicate that STOV metasurfaces are potentially favorable platforms to explore high-dimensional topological physics.

\begin{figure*}[htbp]
\centerline{\includegraphics[width=7in] {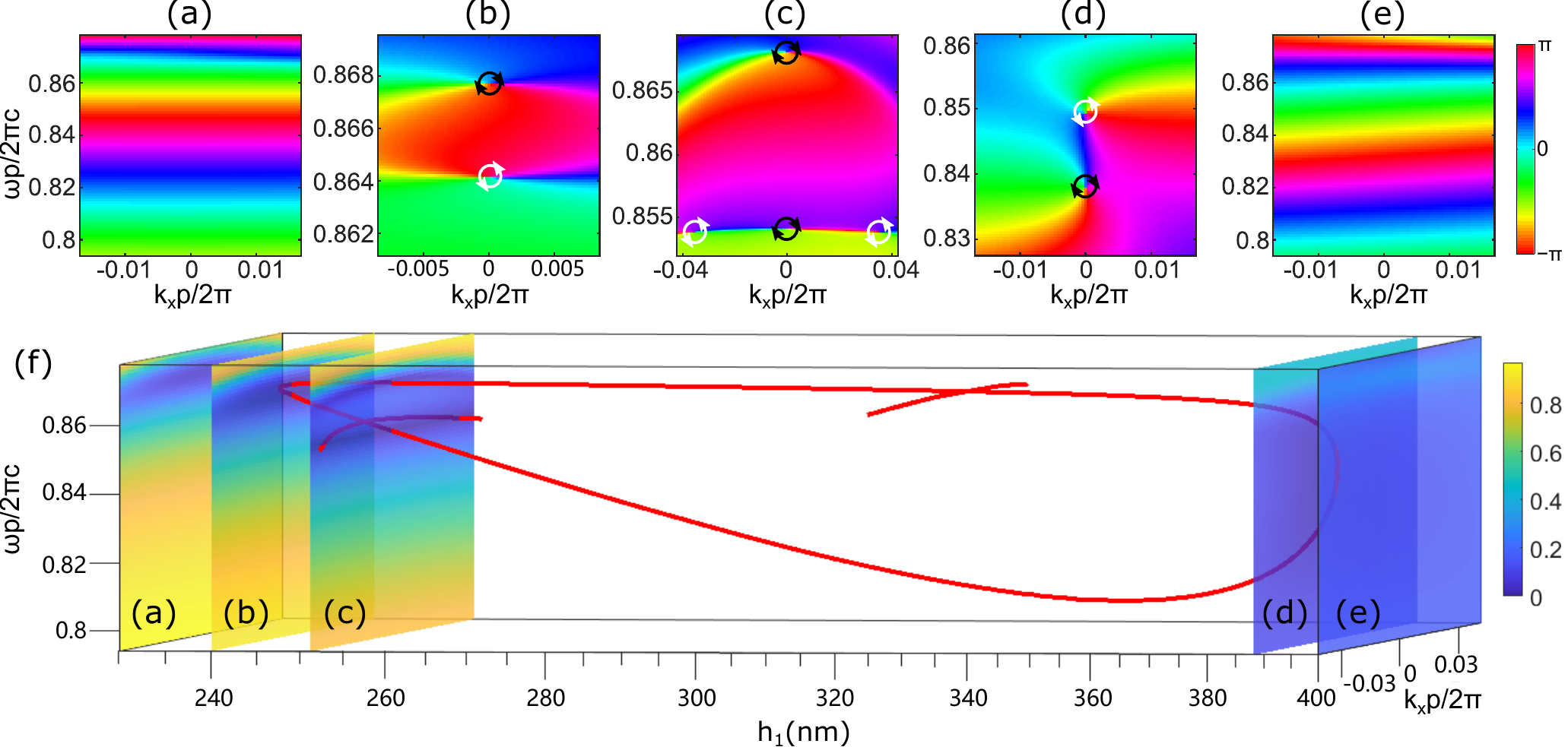}}
\caption{(a-e) The phase distributions of transmission spectrum functions $T$ at ${{h}_{1}}=230$ nm, $240$ nm, $251.3$ nm, $388.3$ nm and $400$ nm, respectively. (f) The nodal line in the three-dimensional synthetic parameter space of the structure. The insets are the amplitude distributions corresponding to (a-e).
 \label{fig:2}}
\end{figure*}

In order to elucidate the topological protection, we begin with briefly discussing the physical motivation of breaking the spatial mirror symmetry for the STOV generation \cite{huang2022spatiotemporal}. We first consider a metasurface consisting of two rods in each period etched on a silicon substrate, with one-dimensional grating structure along $x$ [Fig.~\ref{fig:1}(a)].  Suppose that the metasurface has a spatial mirror symmetry, when a pulse with both Gaussian envelopes in spatial and temporal domains impinges on the structure at normal incidence, due to the mirror symmetry, the phase distribution of the transmitted pulse must also be symmetric about the mirror plane. As a result, there is no phase singularity in the spatiotemporal dimension. Therefore, in order to generate STOVs at normal incidence, it is necessary to break the mirror symmetry of the structure for achieving asymmetric phase modulation and creating a phase singularity. In fact, the spatial-mirror-symmetry breaking can be extended to more than one plane, e.g. by designing a two-dimensional asymmetry grating.

In this way, we next investigate the impact of spatial symmetry breaking, by constructing a synthetic dimension with the geometry parameter $h_1$, the height of the left rod in the unit cell. We analyze the phase modulation of the metasurface by considering an incident plane wave polarized in the $y$ direction ${{E}_{\text{in }}}={{\tilde{s}}_{\text{in }}}\exp \left( i{{k}_{x}}x+i{{k}_{z}}z-i\omega t \right)$, where ${{\tilde{s}}_{\text{in}}}$ is the incident amplitude, ${{k}_{x}}$ and ${{k}_{z}}$ are the wavevector components parallel and perpendicular to the structure interface, respectively. Due to the phase match condition for grating diffraction, in order to simplify the analysis, here we consider the case that only the zeroth-order diffraction light is transmitted with the same wavevector component ${{k}_{x}}$: ${{E}_{\text{trans }}}={{\tilde{s}}_{\text{trans }}}\exp \left( i{{k}_{x}}x+i{{k}_{z}}z-i\omega t \right)$, where ${{\tilde{s}}_{\text{tran}}}$ is the transmitted amplitude. Therefore, the spatiotemporal transformation by the metasurface is defined by a transmission spectrum function $T\left( {{k}_{x}},\omega  \right)={{{{\tilde{s}}}_{\text{tran }}}\left( {{k}_{x}},\omega  \right)}/{{{{\tilde{s}}}_{\text{in }}}\left( {{k}_{x}},\omega  \right)}$.

We expect that due to spatial symmetry breaking, the phase singularities can be illustrated by the phase distribution of $T\left( {{k}_{x}},\omega  \right)$. Indeed, Figs.~\ref{fig:2}(a-e) show the phase distributions of the transmission spectrum functions $T$ by gradually increasing $h_1$, which exhibit different topological features. The detailed geometry of the metasurface is described in Sec. I of the Supplementary Material. We numerically calculate the transmission spectrum function $T$ by the finite-element method using the commercial full-wave software package COMSOL. In the simulation, the material dispersion has been taken into account, and the optical constants of silicon are referred to the experimental data in Ref.~\cite{green2008self}. For $h_1<230$ nm, there is no phase singularity in transmission spectrum function [Fig.~\ref{fig:2}(a)], which corresponds to the trivial case. However, due to spatial mirror symmetry breaking, for ${{h}_{1}}$ around $238.5$ nm, there is a topological phase transition from the trivial case to the topological STOV one. For example, when $h_1=240$ nm, Fig.~\ref{fig:2}(b) shows the two phase singularities are with the topological charges $l=\pm 1$ at $\omega =0.8641\times {2\pi c}/{p}$ and $\omega =0.8678\times {2\pi c}/{p}$. By further increasing $h_1$ over $388.7$ nm, the phase singularities disappear because the impact of the spatial symmetry breaking becomes weak, as the left rod is much taller than the other and dominates the symmetric property of the metasurface. We note that the vortices with opposite handedness are the fundamental `charges' in the ${{k}_{x}}\!-\!\omega $ dimensions. Since they are always `created' or `annihilated' together in pairs of opposite charges, we identify that the total topological charge of these vortices is a conserved quantity.  The change of the phase distribution demonstrates that the spatial symmetry breaking leads to the transition from the trivial cases to the topological STOV one in the ${{k}_{x}}\!-\!\omega $ domain [Fig.~\ref{fig:1}(c)].

In order to illustrate the topological phase transition more clearly, Fig.~\ref{fig:2}(f) shows the nodal lines in the synthetic dimension with the geometry parameter $h_1$. These nodal lines are the intersection of $\operatorname{Re}\left( T \right)=0$ and $\operatorname{Im}\left( T \right)=0$ (cf. Supplementary Material Sec.~II). More importantly, the topological structure is similar to the optical vortices of the scalar field in the real space \cite{nye1974dislocations}. Therefore, when a plane with a fixed value $h_1$ has intersections with the nodal lines in the synthetic dimension, two vortices with opposite handedness appear in the cross-section plane [Fig.~\ref{fig:1}(b)]. In general, the geometry parameter $h_1$ as the order parameter has two critical values ${{c}_{1}}$ and ${{c}_{2}}$, during the transition from the trivial cases to the topological STOV ones, and the nodal line with a ring structure is a topological feature as shown in Fig.~\ref{fig:1}(b).

\begin{figure}[htbp]
\centerline{\includegraphics[width=3.2in] {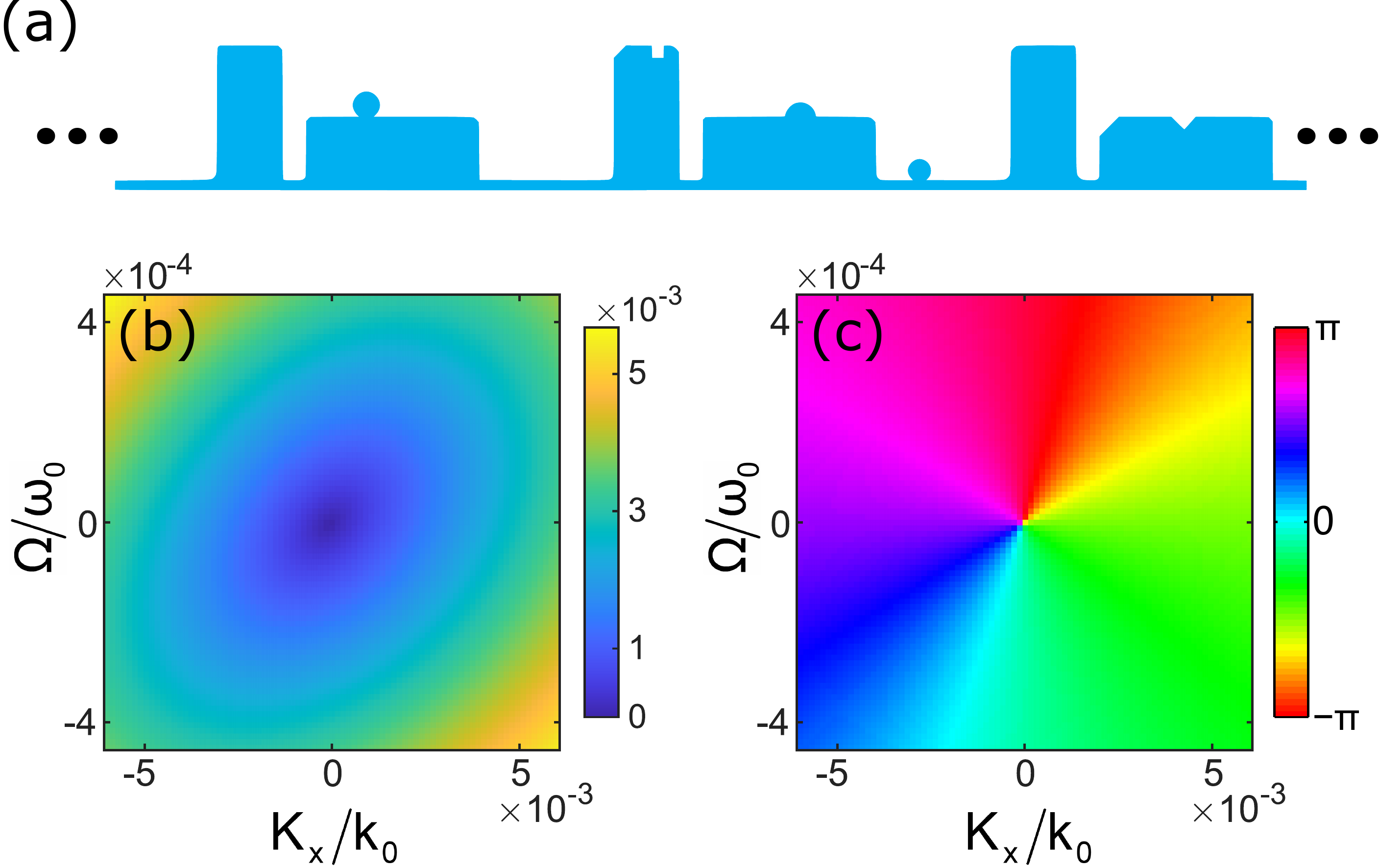}}
\caption{The transmission spectrum function of the STOV generator with random defects. (a) The unit cell of the periodic STOV metasurface with random defects. (b) Amplitude and (c) phase distributions of the transmission spectrum function with respect to ${{K}_{x}}$ and $\Omega$, where the vortex in the ${{k}_{x}}\!-\!\omega $ domain is at ${{\omega }_{0}}= 0.798\times {2\pi c}/{p} $ and $k_{x}^{0}=-0.0288k_0$.
\label{fig:3}}
\end{figure}

We also note that due to the crossovers of nodal lines, for ${{h}_{1}}>248$nm, the ${k_x}\!-\!\omega $ plane has two more interactions with the nodal lines. For example, Fig.~\ref{fig:2}(c) shows the phase distribution of the transmission spectrum function for ${{h}_{1}}=251.3$nm, where two additional vortices are created.  Even when the nodal lines cross over, the total topological charges of these vortices are conserved [Fig.~\ref{fig:2}(c)].  Due to the conservation law and the additional vortices at ${{k}_{x}}\ne 0$ having the same handedness, Fig.~\ref{fig:2}(c) shows that the handedness of the vortex at $\omega =0.854\times {2\pi c}/{p}\;$ and ${{k}_{x}}=0$ has to change from $l=+1$ to $l=-1$ in compassion with Fig.~\ref{fig:2}(b). In the same situation, due to the crossover of nodal lines around ${{h}_{1}}=330$nm, Fig.~\ref{fig:2}(d) shows that the vortex handedness at $\omega =0.849\times {2\pi c}/{p}\;$ and ${{k}_{x}}=0$ changes from $l=-1$ to $l=+1$. The details of the transmission spectrum functions near the crossovers of the nodal lines for ${{h}_{1}}=251.3$nm and ${{h}_{1}}=324.5$nm are described in Sec. III of the Supplementary Material. Finally, the two vortices are annihilated as shown in Fig.~\ref{fig:2}(e) at $h_1=400$ nm.

Next we show that the topological protected vortex in the ${{k}_{x}}\!-\!\omega $ domain can be transferred to the spatiotemporal domain and thus generates the topologically protected STOV. Suppose a vortex with charge number $l$ located at $({{\omega }_{0}},k_{x}^{0})$ in the ${{k}_{x}}\!-\!\omega $ domain, we consider an incident (transmitted) pulse with central frequency ${{\omega }_{0}}$ and transverse wavevector component $k_{x}^{0}$ as ${{E}_\text{in(trans)}}(x,t)={{s}_\text{in(trans)}}(x,t)\exp (ik_{x}^{0}x+ik_{z}^{0}z-i{{\omega }_{0}}t)$. Here ${{s}_\text{in}}(x,t)$ and ${{s}_\text{trans}}(x,t)$ are the spatiotemporal envelope amplitudes of the incident and transmitted fields, respectively. In order to specifically depict the transformation between the incident and transmitted pulses while propagating through the STOV generator, we decompose the incident (transmitted) pulse envelope into a series of plane waves by Fourier transform ${{s}_{\text{in(trans) }}}(x,t)=\iint{{{{\tilde{s}}}_{\text{in(trans) }}}}\left( {{K}_{x}},\Omega  \right)\exp \left( i{{K}_{x}}x-i\Omega t \right)\text{d}{{K}_{x}}\text{d}\Omega $, where ${{K}_{x}}={{k}_{x}}-k_{x}^{0}$ is the wavevector component shifted along $x$ direction, and $\Omega =\omega -{{\omega }_{0}}$ is the sideband angular frequency from the center one ${{\omega }_{0}}$. In the polar coordinate, the expression is converted to ${s_{{\rm{in(trans) }}}}(r,\theta ) = \int\limits_0^\infty  {\int\limits_0^{2\pi } {{{\tilde s}_{{\rm{in(trans) }}}}\left( {\rho ,\chi } \right)\exp \left[ {i\rho r\cos \left( {\chi  + \theta } \right)} \right]\rho {\rm{d}}\rho {\rm{d}}\chi } } $, where $r=\sqrt{{{x}^{2}}+{{t}^{2}}}$, $\theta =\tan{^{-1}}\left( {t}/{x} \right)$ and $\rho =\sqrt{{{K}_{x}}^{2}+{{\Omega }^{2}}}$, $\chi =\tan{^{-1}}\left( {\Omega }/{{{K}_{x}}} \right)$. According to the restriction of the winding number around vortex, within the small enough region, the amplitude of transmission spectrum function can be written as $f\left( \rho  \right)$, and without loss of general $T$ can be expressed as
\begin{equation}
 T(\rho ,\chi )=f\left( \rho  \right)\left( A{{e}^{il\chi }}+B{{e}^{-il\chi }} \right),\label{eq:1}
\end{equation}
where $l$ is the winding number of the vortex, $A$ and $B$ are the two constant parameters, respectively. We note that when the phase singularity is anisotropic, both $A$ and $B$ are nonzero. The overall topological charge is $l$ if $\left| A \right|$ is larger than $\left| B \right|$, and $-l$ otherwise. Suppose that an incident pulse with zero topological charge ${{\tilde{s}}_{in}}\left( \rho  \right)$ impinges on the metasurface. With the transmission spectrum function, the transmitted envelope can be expressed as (see Supplementary Material Sec.~IV for details)
\small{
\begin{equation}
{{s}_{\text{tran }}}(r,\theta )=2\pi {{i}^{l}}\left( A{{e}^{-il\theta }}+B{{e}^{il\theta }} \right)\int_{0}^{\infty }{{{{\!\!\tilde{s}}}_{in}}\left( \rho  \right)f\left( \rho  \right){{J}_{l}}\left( \rho r \right)\rho \text{d}\rho }, \label{eq:2}
\end{equation}}
where ${{J}_{l}}$ is the $l$-th order Bessel function of the first kind. Clearly, Eq. (\ref{eq:2}) shows that the transmitted pulse has a vortex with a nonzero topological charge in the spatiotemporal domain, but with the opposite one in the ${{k}_{x}}\!-\!\omega $ domain.

\begin{figure}[htbp]
\centerline{\includegraphics[width=3.2in] {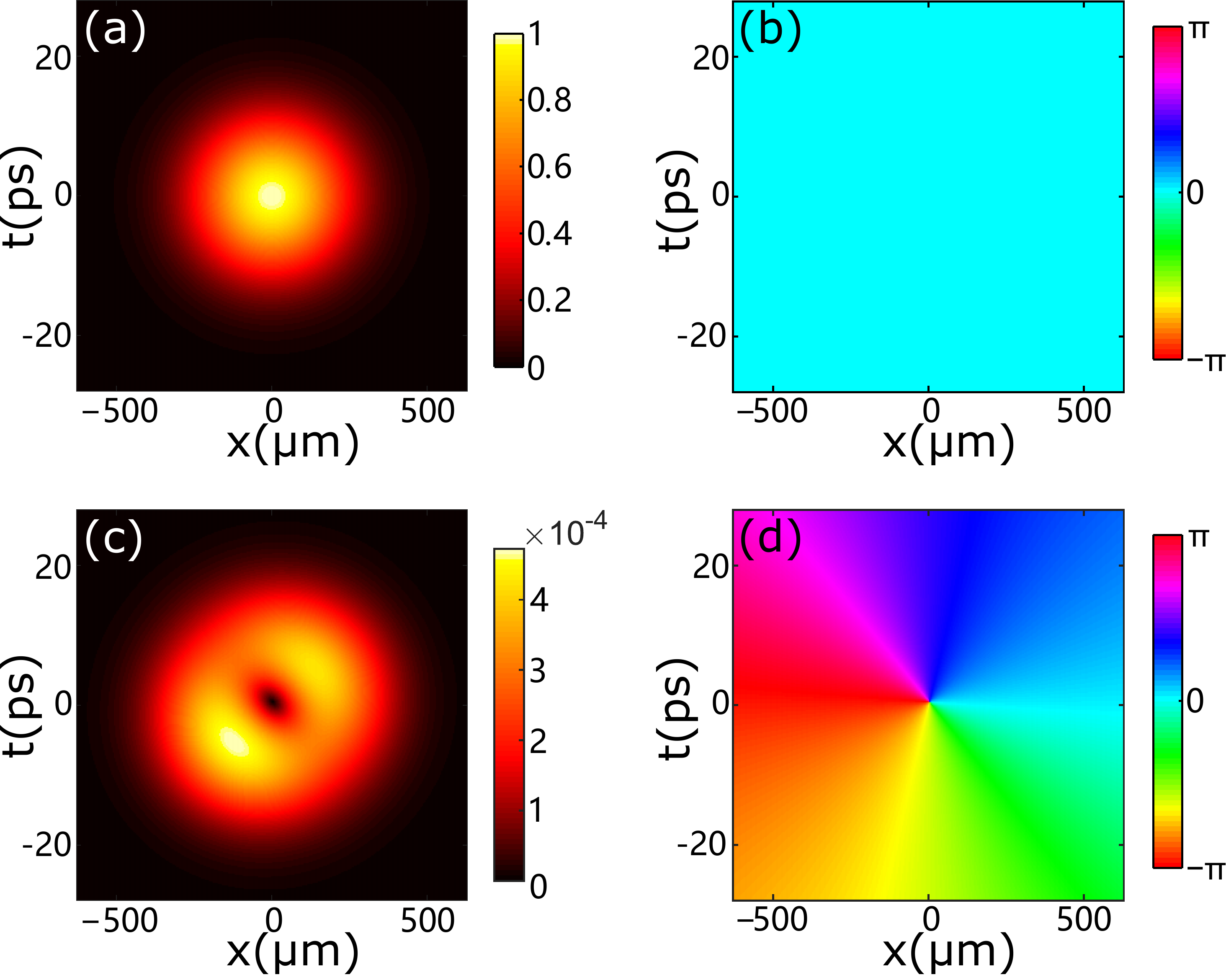}}
\caption{Generation of a STOV by the metasurface with random defects for a Gaussian-enveloped incident pulse. (a) Amplitude and (b) phase distributions of obliquely incident pulse with Gaussian envelopes in both spatial and temporal domains. (c) Amplitude and (d) phase distributions of the transmitted one. The transmitted pulse with transverse OAM has a phase singularity, leading to the zero amplitude at the pulse center.
\label{fig:4}}
\end{figure}

Next we further show that the STOV generation with the metasurface can be topologically protected by the vortices in the ${{k}_{x}}\!-\!\omega $ domain. When the vortices are distanced from the crossovers of nodal lines and the critical points around the topological phase transition, small changes to the metasurface geometry in the real space can be treated as perturbations to the vortex at $({{\omega }_{0}},k_{x}^{0})$. The strength of the topological protection can be evaluated by the distance to its nearest vortex:
\begin{equation}
\Delta =\sqrt{{{({{\omega }_{n}}-{{\omega }_{0}})}^{2}}+c^2{{(k_{x}^{n}-k_{x}^{0})}^{2}}},\label{eq:3}
\end{equation}
where ${{\omega }_{n}}$ and $k_{x}^{n}$ are the frequency and the wavevector component of the nearest vortex. When $\Delta $ is large enough, the vortices with $l=\pm 1$ are stable due to topological protection from the perturbation: Since the continuous structural perturbation only continuously perturbs the contours $\operatorname{Re}\left( T \right)=0$ and $\operatorname{Im}\left( T \right)=0$, the intersection of the two contours still gives vortices in the ${{k}_{x}}\!-\!\omega $ domain. On the other hand, when the vortices are close to the crossovers of nodal lines or the critical points of the transitions, intrinsically, these vortices are vulnerable and cannot be robust to perturbations, and it is meaningless to define the protection strength for these vulnerable vortices.

To demonstrate the robust generation of STOV against structural perturbations and disorders, we randomly generate some defects shown in Fig.~\ref{fig:3}(a), where $h_1=340$ nm and the topological protection strength $\Delta =0.059\times {2\pi c}/{p}$.  The defects are randomly positioned, including partial convexities, concavities, and various kinds of chamfers (the detailed geometry cf. Supplementary Material Sec.~I). We numerically calculate the amplitude and phase distributions of the transmission spectrum function $T$ [Figs.~\ref{fig:3}(b-c)] for the zeroth-order diffraction light. Indeed, even with such perturbations, the phase distribution of $T$ exhibits that the vortex in the ${{k}_{x}}\!-\!\omega $ domain still survives, which leads to a zero amplitude at ${{\omega }_{0}}=0.798\times {2\pi c}/{p}$ and $k_{x}^{0}=-0.0288 k_0$, where $k_0 = {{{\omega }_{0}}}/{c}$. Clearly, it indicates the topologically protected robustness of the STOV metasurface.

Next we simulate an incident pulse with Gaussian envelopes in both spatial and temporal domains to demonstrate the STOV generation. Figures \ref{fig:4}(a) and (b) show the amplitude and phase distributions of the incident pulse envelope, where the beam waist and pulse width are 502 $\mu$m and 22 ps within the ranges of spatial and temporal bandwidths shown in Fig.~\ref{fig:3}. After passing through the metasurface, we simulate the transmitted pulse by the Fourier transform method, with the transmission spectrum function in Fig.~\ref{fig:3}. The amplitude and phase distributions of transmitted pulse envelope are shown in Figs.~\ref{fig:4}(c) and (d), respectively. Indeed, they show that with the topological protection, the transmitted pulse is still a STOV with zero amplitude at the pulse center [Fig.~\ref{fig:4}(c)] and has a phase singularity [Fig.~\ref{fig:4}(d)] in the spatiotemporal domain, which successfully immunizes against the defects in the metasurface.

In summary, we investigate the topological property of STOV metasurface in a synthetic parameter dimension associated with the metasurface geometry. Since STOV metasurfaces are much more compact than the pulse shapers, their robust immunity to fabrication defects paves the way forward generating and modulating the transverse OAM with integrated devices. We also note that here the chosen geometry parameter $h_1$ is not particular, and there are plenty of the other geometry parameters which can realize the STOVs. Therefore, the nodal line discussed in the present paper is transformed to a hyperplane in the high-dimensional synthetic space. It is intriguing to explore the topological structure in such a synthetic space, associated with frequency and wavevector. We believe that there are high-order vortices emerging in the synthetic parameter dimension, which could be important in the applications of STOVs.

The authors acknowledge funding through the National Key Research and Development Program of China (Grant No. 2017YFA0205700), the National Natural Science Foundation of China (NSFC Grants Nos. 12174340 and 91850108), the Open Foundation of the State Key Laboratory of Modern Optical Instrumentation, and the Open Research Program of Key Laboratory of 3D Micro/Nano Fabrication and Characterization of Zhejiang Province.

%

\newpage
\clearpage

\setcounter{section}{0}

\newcommand{\hbAppendixPrefix}{S}
\renewcommand{\thefigure}{\hbAppendixPrefix\arabic{figure}}
\setcounter{figure}{0}

\renewcommand{\thetable}{\hbAppendixPrefix\arabic{table}}
\setcounter{table}{0}
\renewcommand{\theequation}{\hbAppendixPrefix\arabic{equation}}
\setcounter{equation}{0}

\onecolumngrid
\begin{center}
\large{\textbf{
Supplementary Material: Topologically protected generation of spatiotemporal optical vortices }}
\end{center}

\section{The detailed geometry of STOV generator metasurface without and with defects.}
\begin{figure}[H]
\centerline{\includegraphics[width=6in] {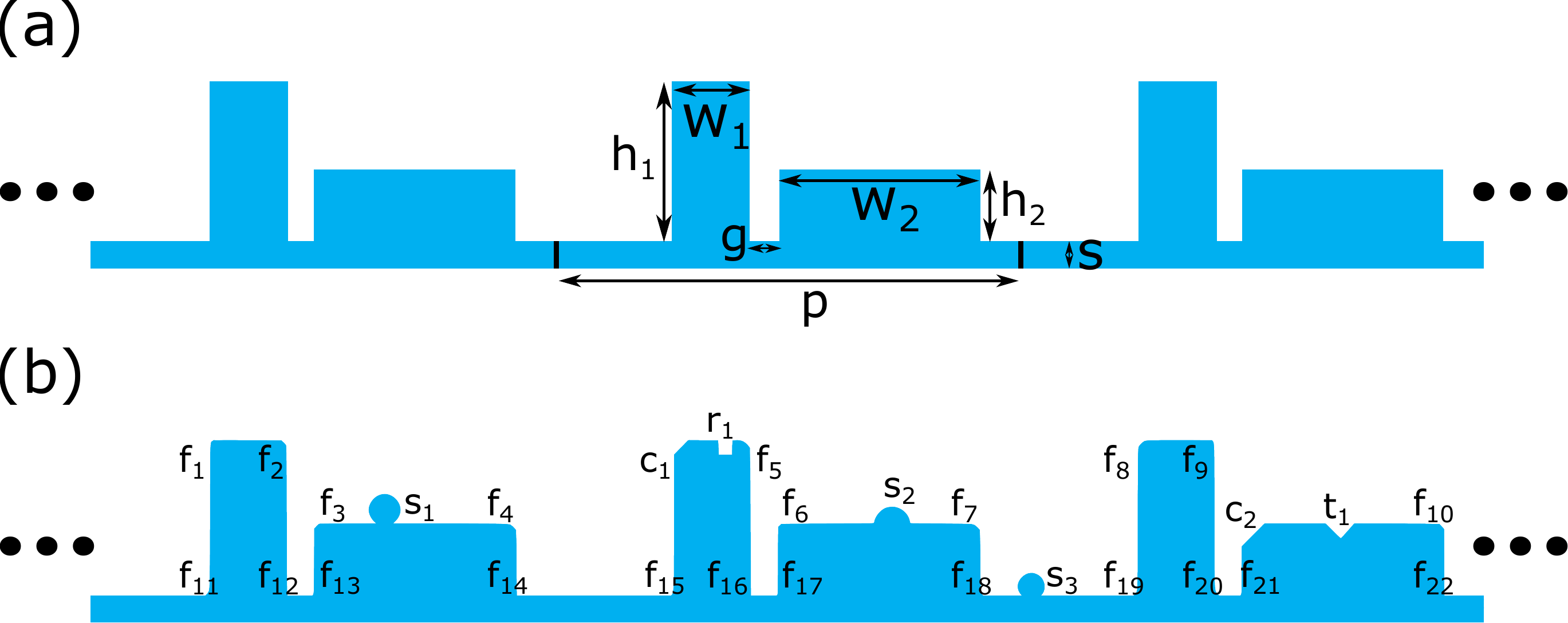}}
\caption{(a) The geometry of STOV generator metasurface.  (b)The unit cell of the periodic STOV generator with random defects [Fig. 3(a)]. $f_1$ to $f_{22}$ correspond to the fillets of rods. $c_1$ and $c_2$ are the chamfers on the rods. $s_1$ to $s_{3}$ are three spherical convexities on the metasurface. $r_1$ and $t_1$ are two concavities with rectangular and triangular shapes.
\label{fig:S1}}
\end{figure}
\begin{figure}[H]
\centerline{\includegraphics[width=5in] {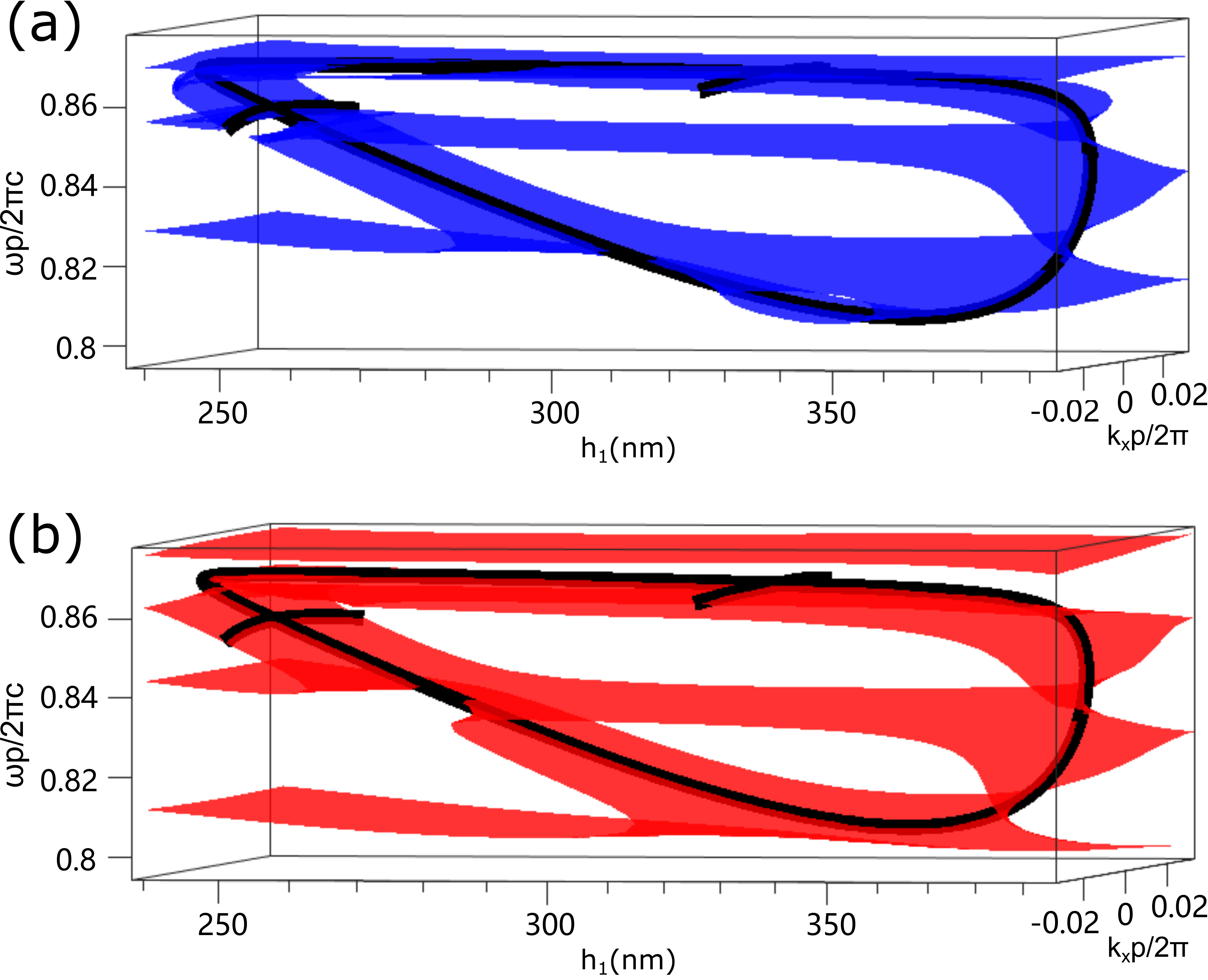}}
\caption{(a) The $\operatorname{Re}\left( T \right)=0$ (blue surface) and (b) $\operatorname{Im}\left( T \right)=0$ (red surface) in the synthetic parameter dimension, respectively. The nodal lines (black) are formed by the intersection set of all the points on both the two surfaces.
\label{fig:S2}}
\end{figure}

As shown in Fig. \ref{fig:S1}(a), the geometry parameters of the metasurface except $h_1$ are the height of the right rod ${{h}_{2}}=160$ nm, the thickness of substrate $s=20$ nm, the gap between two rods $g=64$ nm, the period $p=1000$ nm, the widths of two rods ${{w}_{1}}=160$ nm and ${{w}_{2}}=432$ nm, respectively. The metasurface with random defects is shown in Fig. \ref{fig:S1}(b). The radiuses of fillets on the rods shown as $f_1$ to $f_{22}$ in Fig. \ref{fig:S2} are 5.6 nm, 14.7 nm, 13.4 nm, 19.4 nm, 22.6 nm, 8.3 nm, 20.4 nm, 19.7 nm, 4.9 nm, 15 nm, 28.8 nm, 10.2 nm, 17.6 nm, 6.7 nm, 22.5 nm, 7.7 nm, 15.2 nm, 21 nm, 26.7 nm, 28.8 nm, 16.4 nm, 4.2 nm, respectively. The distances from vertices of the chamfers on the rods labeled as $c_1$ and $c_2$ are 28.9 nm and 45.9 nm. The radius of the first spherical convexity $s_1$ is 32.2 nm. It is tangent to top of the rod and the center of it is 149.8 nm away from the left edge of the rod. The second spherical convexity $s_2$ is with 37-nm radius. The center of it is on top of the rod and 243.9 nm away from the left edge of the rod. The radius of the third spherical convexity $s_3$ is 27.1 nm. It is tangent to the substrate and the center of it is 111.9 nm away from the right edge of the fourth rod counted from left to right. The length and width of the rectangular concavity $r_1$ are 33.5 nm and 31.6 nm, respectively. The distance between the left edge of the concavity and the left edge of the rod is 92 nm. The height and base of the isosceles-triangle shaped concave $t_1$ are 33 nm and 62 nm, respectively, and the distance between the left angle of it and the left edge of the rod is 178 nm. The parameters above are randomly generated, therefore the results in Fig. 3 and Fig. 4 demonstrate the robust topological protection of STOV generator.

\section{The $\textbf{Re}(T)=0$ and $\textbf{Im}(T)=0$ in the synthetic parameter dimension.}

\section{Phase distributions of transmission spectrum functions near the crossovers of the nodal lines.}
{
\begin{figure}[H]
\centerline{\includegraphics[width=5in] {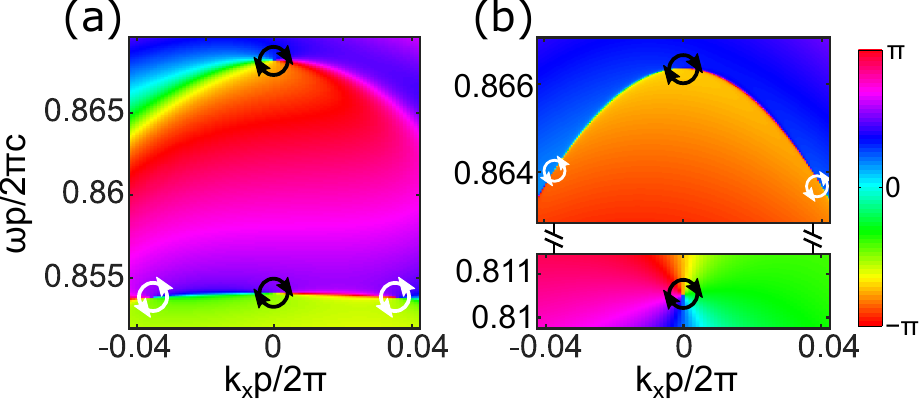}}
\caption{(a-b) The phase distributions of transmission spectrum functions $T$ near the crossovers of the nodal lines for $h_1=251.3$nm and $h_1=324.5$nm, respectively.
\label{fig:S3}}
\end{figure}}

\section{The detailed derivation about Eq. (2).}
Suppose that an incident pulse with zero topological charge ${{\tilde{s}}_{in}}\left( \rho  \right)$ impinges on the metasurface, the transmitted pulse can be expressed as:
\begin{equation}
\begin{aligned}
   {{s}_{trans}}(r,\theta )& = \iint{{{{\tilde{s}}}_{in}}(\rho )f(\rho )\left( A{{e}^{il\chi }}+B{{e}^{-il\chi }} \right)}\cdot {{e}^{i({{k}_{x}}x-\omega t)}}d{{k}_{x}}d\omega  \\
& = \int_{0}^{\infty }{{{{\tilde{s}}}_{in}}(\rho )f(\rho )\rho \int_{0}^{2\pi }{A{{e}^{il\chi }}{{e}^{i(\rho \cos \chi \cdot r\cos \theta -\rho \sin \chi \cdot r\sin \theta )}}+B{{e}^{-il\chi }}{{e}^{i(\rho \cos \chi \cdot r\cos \theta -\rho \sin \chi \cdot r\sin \theta )}}d\chi }}d\rho  \\
 & = \int_{0}^{\infty }{{{{\tilde{s}}}_{in}}(\rho )f(\rho )\rho }\left[ A{{e}^{-il\theta }}\int_{0}^{2\pi }{{{e}^{i[-l(-\chi -\theta )+\rho r\cos (\chi +\theta )]}}d\chi }+B{{e}^{il\theta }}\int_{0}^{2\pi }{{{e}^{i[-l(\chi +\theta )+\rho r\cos (\chi +\theta )]}}d\chi } \right]d\rho .
\end{aligned}
\end{equation}
Taking $\eta =-\chi -\theta +\frac{\pi }{2}$ and $\tau \text{=}\chi +\theta +\frac{\pi }{2}$ , then $d\eta =-d\chi $ and $d\tau =d\chi $, by periodicity, the integral of $d\chi $ becomes:
\begin{equation}
\begin{aligned}
   &A{{e}^{-il\theta }}\int_{0}^{2\pi }{{{e}^{i[-l(-\chi -\theta )+\rho r\cos (\chi +\theta )]}}d\chi }+B{{e}^{il\theta }}\int_{0}^{2\pi }{{{e}^{i[-l(\chi +\theta )+\rho r\cos (\chi +\theta )]}}d\chi } \\
 & = -A{{e}^{-il\theta }}\int_{-\theta +\frac{\pi }{2}}^{-2\pi -\theta +\frac{\pi }{2}}{{{e}^{i[-l\left( \eta -\frac{\pi }{2} \right)+\rho r\cos (\frac{\pi }{2}-\eta )]}}d\eta }+B{{e}^{il\theta }}\int_{\theta +\frac{\pi }{2}}^{2\pi +\theta +\frac{\pi }{2}}{{{e}^{i[-l\left( \tau -\frac{\pi }{2} \right)+\rho r\cos (\tau -\frac{\pi }{2})]}}d\tau } \\
 & = -{{i}^{l}}A{{e}^{-il\theta }}\int_{0}^{-2\pi }{{{e}^{i[-l\eta +\rho r\sin (\eta )]}}d\eta }+{{i}^{l}}B{{e}^{il\theta }}\int_{0}^{2\pi }{{{e}^{i[-l\tau +\rho r\sin (\tau )]}}d\tau } \\
 & = {{i}^{l}}A{{e}^{-il\theta }}\int_{0}^{2\pi }{{{e}^{i[-l\eta +\rho r\sin (\eta )]}}d\eta }+{{i}^{l}}B{{e}^{il\theta }}\int_{0}^{2\pi }{{{e}^{i[-l\tau +\rho r\sin (\tau )]}}d\tau } \\
 & = 2\pi {{i}^{l}}(A{{e}^{-il\theta }}+B{{e}^{il\theta }}){{J}_{l}}(\rho r).
\end{aligned}
\end{equation}
Where ${{J}_{l}}\left( \rho r \right)=\frac{1}{2\pi }\int_{0}^{2\pi }{{{e}^{i\left[ -l\gamma +\rho r\sin (\gamma ) \right]}}d\gamma }$ is the $l$-th order Bessel function of the first kind. Therefore, the transmission spectrum function can be expressed as ${{s}_{trans}}(r,\theta )=2\pi {{i}^{l}}(A{{e}^{-il\theta }}+B{{e}^{il\theta }})\int_{0}^{\infty }{{{{\tilde{s}}}_{in}}(\rho )f(\rho ){{J}_{l}}(\rho r)\rho d\rho }$.

\end{document}